\documentclass[conference]{IEEEtran}
\IEEEoverridecommandlockouts
\usepackage[T1]{fontenc}
\usepackage{cite}
\usepackage{amsmath,amssymb,amsfonts}
\usepackage{booktabs}
\usepackage{bm}
\usepackage{algorithmic}
\usepackage{graphicx}
\usepackage{textcomp}
\usepackage{xcolor}

\usepackage[hyphens]{url}
\usepackage{hyperref}
\usepackage{xurl}
\hypersetup{breaklinks=true}

\def\vzero{{\mathbf{0}}}
\def\vone{{\mathbf{1}}}

\def\vxi{{\boldsymbol{\xi}}}

\def\vb{{\mathbf{b}}}
\def\vc{{\mathbf{c}}}
\def\vd{{\mathbf{d}}}

\def\vf{{\mathbf{f}}}

\def\vp{{\mathbf{p}}}
\def\vq{{\mathbf{q}}}
\def\vr{{\mathbf{r}}}
\def\vs{{\mathbf{s}}}

\def\vv{{\mathbf{v}}}
\def\vw{{\mathbf{w}}}
\def\vx{{\mathbf{x}}}

\def\mA{{\mathbf{A}}}

\def\mM{{\mathbf{M}}}

\def\mT{{\mathbf{T}}}

\begin{document}

\bstctlcite{IEEEexample:BSTcontrol}

\title{Network-Aware Reserve Procurement via Decision-Dependent Adaptive Robust Optimization
\thanks{This work was funded by the Leverhulme International Professorship with grant reference LIP-2020-002 and the Engineering and Physical Sciences Research Council under the grant EP/Y025946/1 (Electric Power Innovation for a Carbon-free Society (EPICS)).}
}

\author{\IEEEauthorblockN{Guillaume Van Caelenberg}
\IEEEauthorblockA{
\textit{Imperial College London}\\
London, UK \\
g.van-caelenberg@imperial.ac.uk}
\and
\IEEEauthorblockN{Akylas Stratigakos}
\IEEEauthorblockA{
\textit{University College London}\\
London, UK \\
a.stratigakos@ucl.ac.uk}
\and
\IEEEauthorblockN{Elina Spyrou}
\IEEEauthorblockA{
\textit{Imperial College London}\\
London, UK \\
evangelia.spyrou@imperial.ac.uk}
}

\maketitle
\begin{abstract}
Network congestion can hinder reserve deliverability, motivating network-aware reserve procurement.
Existing adaptive robust optimization approaches 
overlook the dependence of uncertainty sets on reserve demand curves and renewable curtailment, which is important in systems with high levels of variable renewable energy. We address this using decision-dependent box uncertainty sets, whose bounds depend on reserve demand curves and day-ahead curtailment of variable renewable energy.
We reformulate the problem using a primitive uncertainty set and solve it via column-and-constraint generation.
On a 29-bus Great Britain system, the proposed method is more cost-effective than the inscribed boxes method because it accounts for the potential of reserve activation to both aggravate and relieve network congestion, whereas the inscribed boxes method only captures the former. Cost savings stem from differences in energy and reserve schedules, including the beneficial simultaneous activation of upward and downward reserves for real-time balancing in some scenarios.
\end{abstract}
\section{Introduction}
System operators procure reserves, i.e., spare generation capacity, to manage the limited predictability of load and variable renewable energy (VRE) between the day-ahead (DA) and real-time (RT) scheduling stages \cite{devos2019dynamic}.
DA reserve procurement must consider network constraints to ensure reserve deliverability, i.e., that reserve activation in RT is not hindered by transmission congestion \cite{papavasiliou2022_multi_area}.
Adaptive robust optimization (ARO) models \cite{bertsimas2013adaptive} can account for network constraints and ensure reserve deliverability considering balancing errors that belong to a predefined uncertainty set.
ARO models can be used to help system operators parametrize reserve activation policies \cite{bertsimas2024marginal} or contingency scenarios considered in DA \cite{ vancaelenberg2026_constructing_deployment_scenarios}.

Predefined uncertainty sets assume that reserve requirements are fully met.
In practice, however, this may not be the case. System operators use price-elastic demand curves for reserves to balance reserve procurement costs against RT reliability risks \cite{hogan2013_scarcity_pricing_ordc}.
Refs. \cite{nside2022_co_opt_bal_capacity, madani2026_models_reserve_deliverability} consider a demand curve for reserves and identify deliverable reserve schedules using a novel method, \textrm{inscribed boxes}, that  considers an inner approximation of the RT feasible region.
For ARO models, the use of reserve demand curves means that the uncertainty set depends on the volume of met reserve demand, which is a DA decision.
Another DA scheduling decision that changes the uncertainty set is VRE curtailment, which leads to higher downward reserve needs and lower upward reserve needs.
In both cases, the uncertainty set becomes \textit{decision-dependent}.  

In \cite{zhao2015_variable_resource_dispatch}, an ARO model with a decision-dependent uncertainty set is applied to find the largest ranges of RT VRE injections that the system can accommodate by redispatching its available flexibility.
\cite{wang2017_robust_risk_constrained_uc, shao2017_scuc_flexible_uncertainty_set} optimize the bounds of the wind uncertainty set, and value realizations outside the set via a risk constraint and penalty on expected RT curtailment and load shedding;
however, they do not explicitly define reserve products and schedule wind at its point forecast, disregarding DA curtailment.
In contrast, \cite{chen2023_robust_generation_dispatch, liu2026_two_stage_robust_wind_reserve_decision_dependent} optimize a curtailment threshold that caps RT VRE output and thereby truncates the uncertainty set, instead of requiring higher downward reserves.

This article contributes an ARO formulation for reserve procurement with a decision-dependent box uncertainty set, considering dependencies induced by reserve demand curves and DA VRE curtailment.
We solve this problem using two methods for a $29$-bus system resembling Great Britain (GB): \emph{(i)} column-and-constraint generation (CCG), yielding a reserve product with deployment forecast error scenarios, an approach currently used by some system operators \cite{vancaelenberg2026_constructing_deployment_scenarios}; and \emph{(ii)} inscribed boxes \cite{madani2026_models_reserve_deliverability}, which consider an inner approximation of the RT feasible region.
To solve the problem with CCG, we first reformulate the decision-dependent uncertainty set into a decision-independent one \cite{zhang2017_roc_adjustable_uncertainty_sets}; this yields a two-stage ARO problem with full recourse, i.e., that RT recourse actions can fully adapt to forecast errors\cite{nside2022_co_opt_bal_capacity}.

In summary, this article contributes to the literature:
\begin{itemize}
    \item {a method for identifying DA deployment forecast error scenarios for scheduling problems that allow VRE curtailment and procure reserves using demand curves}
    \item {an extension of the inscribed boxes method for VRE curtailment and co-optimization of energy and reserves}
    \item {a comparison of the inscribed boxes method to reserve procurement with deployment scenarios, identified using CCG, for a case study of a GB-like power system.}
\end{itemize}
Numerical simulations on a $29$-bus GB-like system show that reserve procurement with deployment scenarios is  on average more cost-effective than the inscribed boxes method as it considers both positive and negative impacts of reserve activation on network congestion, whereas the inscribed boxes only consider how RT recourse actions can aggravate congestion.
Our analysis reveals that in this case study the cost savings arise mostly due to different spatial distributions of energy and reserve schedule.
The CCG method even procures more upward reserve than demanded in about $7$\% of the hours when counter-trading, i.e., the simultaneous activation of upward and downward reserves at different nodes, helps it access less expensive electricity generation in DA.
The inscribed boxes method becomes relatively less costly when the penalty for unmet downward reserve demand decreases, but this comes at the expense of increased RT curtailment.

The rest of this paper is structured as follows.
Section \ref{problem} formulates the reserve procurement problem, Section \ref{solution_strategies} presents the two methods, Section \ref{experiments} discusses the numerical experiments, and Section \ref{conclusions} provides conclusions.
\section{Problem Formulation} \label{problem}
In this section, we introduce the VRE uncertainty modeling, formulate the ARO with decision-dependent uncertainty, and reformulate the problem via a primitive uncertainty set.
\paragraph*{Notation}
We consider a power system with a set of nodes $\mathcal{N}$, AC lines $\mathcal{L}$, and generators $\mathcal{G}$.
We consider a single-interval scheduling model that meets load $\mathbf{d} \in \mathbb{R}_+^{|\mathcal{N}|}$.
Define $[\vx]^+ = \max\{\vzero, \vx\}$ (element-wise), reformulated via epigraph: $\vx' = [\vx]^+, \vx' \geq \vx, \vx' \geq \vzero$ \cite{boyd2004convex}.
\subsection{VRE Uncertainty Modeling}
The uncertainty arises from RT forecast errors
$\vxi\in\mathbb{R}^{|\mathcal{N}|}$ in VRE production.
The errors are realized production minus DA point forecast ($\hat{\vv} \in \mathbb{R}_+^{|\mathcal{N}|}$). In DA, a predicted forecast error range is also available: $\vxi\in[\underline{\vxi}, \overline{\vxi}]$ and VRE is curtailed by $\vs \in \mathbb{R}_+^{|\mathcal{N}|}$.
Reserves manage the \textit{balancing error} given by
\begin{equation}\label{eq:balance_error}
\boldsymbol{\xi}^{\textrm{b}} = \boldsymbol{\xi} + \vs,
\end{equation}
which, evidently, belongs to the interval $[\underline{\vxi}+\vs, \overline{\vxi}+\vs]$. 
Upward and downward reserves help manage negative  and positive balancing errors, respectively.
Hence, the upper and lower bounds of \eqref{eq:balance_error} help us determine the demand for downward and upward reserves when they are respectively positive and negative.
We allow for demand to be unmet by applying a penalty ($\mathbf{b}^{\textrm{u}}, \mathbf{b}^{\textrm{d}} \in \mathbb{R}_+^{|\mathcal{N}|}$)
to variables that track reserve scarcity  $\mathbf{q}^{\textrm{u}}, \mathbf{q}^{\textrm{d}} \in \mathbb{R}_+^{|\mathcal{N}|}$.
These variables are limited by the demand for upward and downward reserves as shown below:
\begin{align} \label{demand_curve_bounds}
\mathbf{q}^{\textrm{u}} \leq [-\underline{\boldsymbol{\xi}} - \vs]^{+}, \quad 
\mathbf{q}^{\textrm{d}} \leq \overline{\boldsymbol{\xi}} + \vs.
\end{align}
Note that the bound on scarcity of upward reserves does not coincide with the lower bound of the balancing error when DA curtailment exceeds $\underline{\boldsymbol{\xi}}$, as in that case the balancing error will always be positive, and no upward reserves are needed.
Next, we define an uncertainty set for the balancing error
\begin{equation} \label{ddu_set}
\mathcal{U}(\vs,  \vq^{\textrm{u}}, \vq^{\textrm{d}})=\{ \boldsymbol{\xi}^{\textrm{b}}: -[-\underline{\boldsymbol{\xi}} - \vs]^{+} + \mathbf{q}^{\textrm{u}} \leq \boldsymbol{\xi}^{\textrm{b}} \leq \overline{\vxi} + \vs - \mathbf{q}^{\textrm{d}} \},
\end{equation}
where as part of the lower bound we consider $-[-\underline{\boldsymbol{\xi}} - \vs]^{+}$ and not the lower bound implied by \eqref{eq:balance_error}.
We make this choice as we should allow the downward reserve to be fully unmet via $\vq^{\textrm{d}}$, hence the set should always include $\vzero$.
We linearize $[-\underline{\boldsymbol{\xi}} - \vs]^{+}$ via its epigraph, which in this case is tight at the optimum as a non-tight relaxation would enlarge the uncertainty set, which cannot reduce cost since robust feasibility is preserved on any subset of the uncertainty set \cite{bertsimas2020_rob_and_adapt_opt}, unless it is offset by additional upward reserve scarcity $\vq^{\textrm{u}}$, which would increase costs as the penalty is non-negative.

\subsection{Decision-Dependent Adaptive Robust Optimization}
We formulate the following full-recourse ARO problem:
\begin{subequations} \label{aro_ddu}
\begin{alignat}{2}
\min_{} \; & \rlap{$\displaystyle \mathbf{c}^\top\mathbf{p} + (\mathbf{c}^{\textrm{u}})^\top\mathbf{r}^{\textrm{u}} + (\mathbf{c}^{\textrm{d}})^\top\mathbf{r}^{\textrm{d}} + (\vb^{\textrm{u}})^\top\vq^{\textrm{u}} + (\vb^{\textrm{d}})^\top\vq^{\textrm{d}},$} \label{aro_ddu_obj} \\
\text{s.t.} & \; \eqref{demand_curve_bounds}, \label{aro_ddu_c1} \\
& \; \vone^{\top}\vp + \vone^{\top}(\hat{\vv}-\vs) = \vone^{\top}\vd, \label{aro_ddu_c2} \\
& \; \vf^{\textrm{e}} = \mM(\mA\vp + \hat{\vv}-\vs - \vd), \label{aro_ddu_c3} \\
& \; -\overline{\vf} \leq \vf^{\textrm{e}} \leq \overline{\vf}, \label{aro_ddu_c4} \\
& \; \vp + \vr^{\textrm{u}} \leq \overline{\vp}, \label{aro_ddu_c5} \\
& \; \vp - \vr^{\textrm{d}} \geq \underline{\vp}, \label{aro_ddu_c6} \\
& \; \vone^\top\vp^{\textrm{rt}}(\vxi^{\textrm{b}}) + \vone^\top\vxi^{\textrm{b}} = 0, &\quad& \forall \vxi^{\textrm{b}} \in \mathcal{U}, \hspace{2.5em} \label{aro_ddu_c7} \\
& \; \mM(\mA\vp^{\textrm{rt}}(\vxi^{\textrm{b}}) + \vxi^{\textrm{b}}) \leq \overline{\vf} - \vf^{\textrm{e}}, && \forall \vxi^{\textrm{b}} \in \mathcal{U}, \label{aro_ddu_c8} \\
& \; -\mM(\mA\vp^{\textrm{rt}}(\vxi^{\textrm{b}}) + \vxi^{\textrm{b}}) \leq \overline{\vf} + \vf^{\textrm{e}}, && \forall \vxi^{\textrm{b}} \in \mathcal{U}, \label{aro_ddu_c9} \\
& \; -\vr^{\textrm{d}} \leq \vp^{\textrm{rt}}(\vxi^{\textrm{b}}) \leq \vr^{\textrm{u}}, && \forall \vxi^{\textrm{b}} \in \mathcal{U}. \label{aro_ddu_c10}
\end{alignat}
\end{subequations}
Problem \eqref{aro_ddu} calculates the least-cost
energy and reserve schedule $\vp, \vr^{\textrm{u}},  \vr^{\textrm{d}} \in \mathbb{R}^{|\mathcal{G}|}$, with linear costs $\vc, \vc^{\textrm{u}}, \vc^{\textrm{d}}$, including unmet reserve demand $\vq^{\textrm{u}},\vq^{\textrm{d}}$, penalized at $\vb^{\textrm{u}}, \vb^{\textrm{d}}$.
Constraint \eqref{aro_ddu_c1} defines the bounds on unmet reserve demand.
Constraints \eqref{aro_ddu_c2}-\eqref{aro_ddu_c6} impose DA power balance, transmission feasibility, and generator minimum and maximum production capacity, where $\overline{\vf} \in \mathbb{R}_+^{|\mathcal{L}|}$ is the line capacity, $\overline{\vp}, \underline{\vp} \in \mathbb{R}_+^{|\mathcal{G}|}$ are the generator technical limits, $\mM \in {\mathbb{R}}^{|\mathcal{L}| \times |\mathcal{N}|}$ is the Power Transfer Distribution Factors (PTDF) matrix and $\mA \in {\mathbb{R}}^{|\mathcal{N}| \times |\mathcal{G}|}$ the node incidence matrix.
Constraints \eqref{aro_ddu_c7}-\eqref{aro_ddu_c10} impose RT power balance and transmission feasibility for all $\vxi^{\textrm{b}} \in \mathcal{U}$ via reserve activation $\vp^{\textrm{rt}}$, bounded by the DA reserve schedule.
\subsection{Reformulation via Primitive Uncertainty Set}
Problem \eqref{aro_ddu} is a two-stage ARO problem with full recourse and \textit{decision-dependent} uncertainty, as $\mathcal{U}$ depends on $\vs, \vq^{\textrm{u}}, \vq^{\textrm{d}}$.
Following \cite{zhang2017_roc_adjustable_uncertainty_sets}, we express the balancing error as
\begin{align} 
\vxi^{\textrm{b}} = \vw \odot (-[-\underline{\boldsymbol{\xi}} - \vs]^{+} + \vq^{\textrm{u}}) + (\vone - \vw) \odot (\overline{\vxi} + \vs - \vq^{\textrm{d}}),
\label{primitive_map}
\end{align}
where $\vw \in \mathcal{W} = [0,1]^{|\mathcal{N}|}$ is an auxiliary variable and $\mathcal{W}$ denotes a primitive uncertainty set.
We use \eqref{primitive_map} to replace \eqref{aro_ddu_c7}-\eqref{aro_ddu_c10} with \eqref{aro_diu_c2} and write the ARO problem with a \textit{decision-independent} uncertainty set $\mathcal{W}$: 
\begin{subequations} \label{aro_diu}
\begin{align}
\min_{} \; & \eqref{aro_ddu_obj}, \label{aro_diu_obj} \\
\text{s.t.} & \; \eqref{demand_curve_bounds}, \eqref{primitive_map}, \eqref{aro_ddu_c2}-\eqref{aro_ddu_c6}, \label{aro_diu_c7} \\
& \; \eqref{aro_ddu_c7}-\eqref{aro_ddu_c10}, \quad \forall \vw \in \mathcal{W}. \label{aro_diu_c2}
\end{align}
\end{subequations}
\section{Solution strategies} \label{solution_strategies}
We present two solution strategies for \eqref{aro_diu}: column-and-constraint generation and the inscribed boxes.
\subsection{Column-and-Constraint Generation}
Since $\mathcal{W}$ is a box set, we reformulate \eqref{aro_diu} by considering the set of vertices $\mathcal{V}^{\textrm{w}} = \{0,1\}^{|\mathcal{N}|}$, which yields
\begin{subequations} \label{aro_vertices}
\begin{align}
\min_{} \; & \eqref{aro_ddu_obj}, \label{aro_vertices_obj} \\
\text{s.t.} & \; \eqref{demand_curve_bounds}, \eqref{primitive_map}, \eqref{aro_ddu_c2}-\eqref{aro_ddu_c6}, \\
& \; \eqref{aro_ddu_c7}-\eqref{aro_ddu_c10}, \quad \forall \vw \in \mathcal{V}^{\textrm{w}}. \label{aro_vertices_rt}
\end{align}
\end{subequations}
Problem \eqref{aro_vertices} can be solved with the CCG algorithm, yielding a reserve product with deployment scenarios \cite{vancaelenberg2026_constructing_deployment_scenarios}.
Although \eqref{primitive_map} appears bilinear in the uncertain parameter $\vw$ and first-stage decisions $\vs,  \vq^{\textrm{u}}, \vq^{\textrm{d}}$, vertex enumeration \eqref{aro_vertices_rt} overcomes this.

\subsection{Inscribed Boxes}
The inscribed boxes method \cite{madani2026_models_reserve_deliverability}  inner approximates the feasible region of the RT balancing constraints in \eqref{aro_vertices} by replacing \eqref{aro_vertices_rt} with a limited number of constraints that are sufficient for reserve deliverability, and, extending (UD-1)-(UD-4) from \cite{madani2026_models_reserve_deliverability} to consider VRE curtailment and co-optimization of energy and reserve procurement, reads as
\begin{subequations} \label{ibox}
\begin{align}
\min_{} \; & \eqref{aro_ddu_obj}, \label{ibox_obj} \\
\text{s.t.} & \; \eqref{aro_ddu_c1}-\eqref{aro_ddu_c6}, \\
& \; \mA\vr^{\textrm{u}} - ([-\underline{\boldsymbol{\xi}} - \vs]^{+} - \vq^{\textrm{u}}) = \mT^\top(\vf^{\textrm{u}} + \vf^{\textrm{ud}}), \label{ibox_bal_u} \\
& \; \mA\vr^{\textrm{d}} - (\overline{\vxi} + \vs - \vq^{\textrm{d}}) = -\mT^\top(\vf^{\textrm{d}} + \vf^{\textrm{ud}}), \label{ibox_bal_d} \\
& \; [\mM\mT^\top]^+ (\vf^{\textrm{u}} + \vf^{\textrm{d}} + \vf^{\textrm{ud}}) \leq \overline{\vf} - \vf^{\textrm{e}}, \label{ibox_del_u} \\
& \; [-\mM\mT^\top]^+ (\vf^{\textrm{u}} + \vf^{\textrm{d}} + \vf^{\textrm{ud}}) \leq \overline{\vf} + \vf^{\textrm{e}}. \label{ibox_del_d}
\end{align}
\end{subequations}
In \eqref{ibox}, the nodal net reserve positions (supply minus requirement, the latter given by the bounds of $\mathcal{U}$) are decomposed into \textit{bilateral reserve exchanges}, where an exchange from $m$ to $n$ can be interpreted as reserve procured at $m$ covering part of the requirement of $n$.
For every ordered pair $m \neq n$, $\vf^{\textrm{u}}, \vf^{\textrm{d}}, \vf^{\textrm{ud}} \in \mathbb{R}_+^{|\mathcal{N}|(|\mathcal{N}|-1)}$ denote the exchanges of upward reserve from $m$ to $n$, of downward reserve from $n$ to $m$, and of both jointly (capturing imbalance netting, i.e., simultaneous upward and downward requirements offsetting each other), all creating a power flow from $m$ to $n$ when activated, and $\mT$ is the exchange incidence matrix, whose row $(m,n)$ has $+1$ in column $m$ and $-1$ in column $n$.
Constraints \eqref{ibox_bal_u}-\eqref{ibox_bal_d} balance the net reserve exports of each node, and \eqref{ibox_del_u}-\eqref{ibox_del_d} limit the resulting flows for each line and direction to the capacity that is not used by the DA flows.
These constraints ignore counter-flows because the $[ \; ]^+$ operator makes all left-hand-side terms non-negative.
Since $\vone^\top\mT^\top = \vzero^\top$, summing \eqref{ibox_bal_u}-\eqref{ibox_bal_d} over all nodes shows that the total procured upward and downward reserve equals the total upward and downward requirement implied by the box size, $\vone^\top([-\underline{\boldsymbol{\xi}} - \vs]^{+} - \vq^{\textrm{u}})$ and $\vone^\top(\overline{\vxi} + \vs - \vq^{\textrm{d}})$.

\section{Numerical experiments} \label{experiments}
\subsection{Test System and Experimental Design}
\begin{figure}[!ht]
\centerline{\includegraphics[width = 0.8\columnwidth]{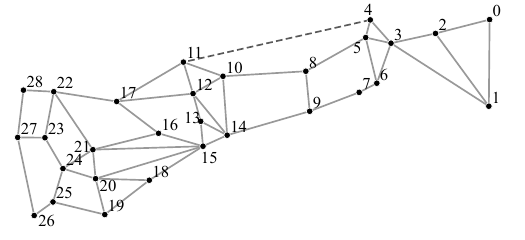}}
\caption{$29$-bus GB power system, with AC (solid) and DC (dashed) lines.
\vspace{-2mm}
} \label{gb_grid}
\end{figure}

We simulate a $29$-bus GB power system (Fig. \ref{gb_grid}) at hourly granularity in $2023$, using the network, load and generation data (thermal, hydro, solar) from PyPSA-GB \cite{lyden2024_pypsagb}.
We assume that only wind has forecast errors and construct the bounds [$\underline{\vxi},\overline{\vxi}$] via Quantile Regression Forests \cite{meinshausen2006_quant_regr_forest}, trained with $2022$ DA point forecasts of NESO \cite{neso_da_wind_forecast} and ERA5 re-analysis actuals (from PyPSA-GB), and consider symmetric $95$\% prediction intervals per node.
The prediction intervals have an average out-of-sample coverage of $93.7$\% across nodes, ranging from a minimum of $90.5$\% to a maximum of $97.3$\%.
We set the upward and downward demand curve penalties to $5000$ £/MW.
The system has one DC line, modeled as an equal injection/withdrawal at the receiving/sending node.
The CCG master problem is warm-started with extreme scenarios ($\vw=\vzero,\vone$).
The CCG subproblem is solved by first applying strong duality and then invoking Gurobi's bilinear solver with relative and absolute MIP gaps of $1e-4$ and $1e-10$, respectively (alternatively, the alternating direction method heuristic can be used) \cite{vancaelenberg2026_constructing_deployment_scenarios}.
\subsection{Results}
\subsubsection{Annual results}
\begin{figure}[!ht]
\centerline{\includegraphics[width = \columnwidth]{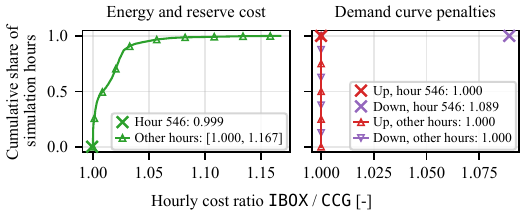}}
\vspace{-2mm}
\caption{Relative value of objective function terms of \texttt{IBOX} w.r.t. \texttt{CCG}.
}
\label{cost_ratio}
\end{figure}
The objective function consists of DA energy costs, DA reserve costs, and penalties for unmet reserve demand. 
In Fig. \ref{cost_ratio},  we plot the cumulative probability function of the \texttt{IBOX} vs. \texttt{CCG} ratio of the energy-reserve cost and penalties for unmet reserve demand. 
Given that the \texttt{IBOX} assumes a smaller feasible RT region than the \texttt{CCG}, the value of the objective function of the \texttt{IBOX} is expected to be higher.  
In this case, for all but one hour (hour $546$), the two methods meet the same level of reserve demand, but the energy-plus-reserve costs of the \texttt{IBOX} method are up to $17$\% higher.

\begin{table}[!ht]
\caption{Average DA cost breakdown (k£/h).}
\label{da_cost}
\vspace{-2mm}
\centering
\begin{tabular}{@{}l|cccc@{}}
\toprule
& Energy & Reserve & Upward penalties & Downward penalties \\
\midrule
\texttt{CCG}  & 1 069.14 & 2.44 & 0.06 & 5 243.73 \\
\texttt{IBOX} & 1 080.88 & 2.24 & 0.06 & 5 243.80 \\
\bottomrule
\end{tabular}
\end{table}

\begin{figure}[!ht]
\centerline{\includegraphics[width = \columnwidth]{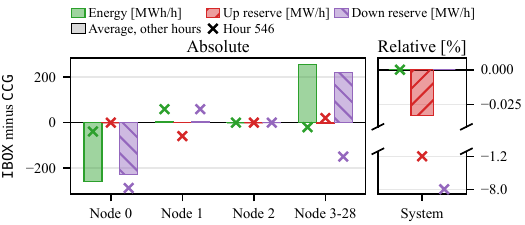}}
\vspace{-2mm}
\caption{Left: average absolute difference in spatial shift energy-reserve schedule. 
Right: average relative difference at the system.
}
\label{zonal_shift}
\end{figure}

Table \ref{da_cost} summarizes the average DA cost per method.
The main cost difference between the methods lies in energy costs.
The higher energy costs under \texttt{IBOX} are largely explained by the shift in the generation mix.
As Fig. \ref{zonal_shift} shows, compared to \texttt{CCG}, electricity generation shifts from node $0$, where the primary generators are wind and hydro, to nodes $3-28$, where more electricity is generated by more expensive CCGT units under \texttt{IBOX}.
This change in the generation mix is due to congestion along the line $0-2$.
To keep transmission capacity available for delivering upward reserves to nodes $0-2$, the \texttt{IBOX} proactively reduces the amount of energy exported from node $0$.
In \texttt{IBOX}, the congestion relief due to lower production at node $0$ can be captured in the energy-related flow, but not in the reserve-induced flow.
In contrast, the \texttt{CCG} accurately captures the congestion relief of lower generation at node $0$, even when it is related to the dispatch of downward reserve.

In  $611$ hours, the \texttt{CCG} even finds it beneficial to procure  $0.03$\%-$3.04$\% more upward reserves than demanded, which equals an average over all hours of $0.03$\% (see Fig. \ref{zonal_shift} right panel).
The over-procurement occurs because the \texttt{CCG} is aware that simultaneous activation of upward and downward reserves (also called counter-trading) can help manage transmission limits.
In $604$ hours, counter-trading is beneficial for meeting the upward reserve needs of node $2$ when line $0-2$ is congested.
Covering a $1$MW upward need with $1$MW upward reserves from nodes $1$/$3$–$28$ overloads line $0$–$2$ as node $2$ has a larger negative PTDF toward that line than nodes $1$/$3$–$28$.
However, a counter-trade of downward reserve from node $0$ with upward reserve from nodes $1$/$3$–$28$ can offset this overload, and the required counter-trade volume only depends on the relative differences of PTDFs.
This beneficial impact of counter-trading cannot be captured by \texttt{IBOX} because the inner approximation only considers congestion-aggravating impacts of reserve deployment.
In this system, activation of upward reserve at nodes other than $0$ to balance upward needs at node $2$ always aggravates congestion of line $0-2$.

As a result, the \texttt{IBOX} method needs to keep more transmission capacity for reserve deliverability in line $0-2$, as evidenced by the higher on average  transmission reserve margin for that line under \texttt{IBOX} in Table \ref{tab_congested_lines}.
Also, in $2$ of the $604$ hours, in addition to line $0-2$, line $20-21$ is congested, and \texttt{CCG} finds it beneficial to simultaneously activate downward reserve at node $19$ and upward reserve at node $28$, which both offset congestion on line $20-21$.
In the remaining $7$ of the $611$ hours (of which hour $546$ is one), line $0-2$ is also congested, but the counter-trading is beneficial for upward reserve needs of nodes $3-28$ covered by upward reserve at node $1$; counter-trading downward reserve at node $0$ with upward reserve at node $2$ offsets the induced congestion.

\begin{table}[!ht]
\caption{Average transmission reserve margin [\% line capacity] and congestion rent [k£/h] in binding hours. In brackets: share of hours in which line is congested in \texttt{CCG}/\texttt{IBOX} [\%].}
\label{tab_congested_lines}
\vspace{-2mm}
\centering
\footnotesize
\setlength{\tabcolsep}{3pt}
\begin{tabular}{@{}l|cc|cc|cc@{}}
\toprule
& \multicolumn{2}{c|}{Line 0-2 (95.9/95.9)} & \multicolumn{2}{c|}{Line 8-10 (0.3/0.3)} & \multicolumn{2}{c}{Line 20-21 (1.4/2.9)} \\
\cmidrule(lr){2-3} \cmidrule(lr){4-5} \cmidrule(l){6-7}
& $\overline{f} - f^{\textrm{e}}$ & Cong. rent & $\overline{f} - f^{\textrm{e}}$ & Cong. rent & $\overline{f} + f^{\textrm{e}}$ & Cong. rent \\
\midrule
\texttt{CCG}  & 7.1   & 5 678.4 & 4.9  & 3.9 & 1.3  & 2.7 \\
\texttt{IBOX} & 32.4  & 5 673.0 & 0.7  & 6.1 & 0.8  & 4.2 \\
\bottomrule
\end{tabular}
\vspace{1mm}
\parbox{\columnwidth}{\footnotesize Note: $\overline{f}-f^{\textrm{e}}$ and $\overline{f}+f^{\textrm{e}}$ are the right-hand sides of \eqref{aro_ddu_c8}-\eqref{aro_ddu_c9}, which limit the reserve-induced flow over the line. We report only one per line, as the direction of congestion is the same in all binding hours.}
\end{table}

\subsubsection{Changes in reserve schedules}
Although the change in energy schedule (Fig. \ref{zonal_shift}) helps explain a large share of the increased energy costs in \texttt{IBOX}, a change in energy schedule is not necessary in all hours. 
During $1092$ hours, the two methods have the same energy schedule but different spatial distributions of reserves, with a lower reserve cost for the \texttt{CCG}.
We further discuss this change by analyzing the results for two clusters of hours, with changes in the reserve schedule.

\paragraph{Spatial shift in upward reserve schedule}

There is a single hour (hour $1976$) in which the schedule for energy, downward reserve, and transmission flows (only line $20-21$ is congested) is the same for both methods, but \texttt{IBOX} shifts $115$MW upward reserve from node $19$ to node $14$, at an upward reserve cost increase of $5.2$\%, compared to \texttt{CCG}.
This shift occurs because \texttt{IBOX} ignores flow relief from reserve activations and forecast errors, whereas CCG exploits these effects to procure cheaper reserves despite their adverse impact on line $20$–$21$, offset by forecast-error-induced flows.

\paragraph{Spatial shift in downward reserve schedule}

In $1091$ hours (low-wind hours without curtailment and unmet reserve demand), \texttt{IBOX} and \texttt{CCG} have the same energy, upward reserve and transmission schedule (only line $0$–$2$ is binding in RT for both methods) but place the downward reserve at different nodes.
Balancing RT wind over-production in nodes $0$ and $1$ with downward reserve in nodes $3$–$28$ loads line $0$–$2$, whereas downward reserve at node $0$ does not.
\texttt{CCG} nets the resulting flows, whereas \texttt{IBOX} ignores counter-flows and shifts on average $192$MW ($4.1$\%) of downward reserve from nodes $3$–$28$ to nodes $0$ and $1$, which reduces the flow on line $0$–$2$ by an average of $47$MW, at a downward reserve cost increase of on average $14.3$\%, compared to \texttt{CCG}.

\subsubsection{VRE curtailment}
The GB system has concentrated VRE production behind transmission constraints, making DA curtailment essential for ensuring nodal supply-demand balance.
For instance, DA curtailment occurs in $70.97$\% and $20.81$\% of the hours at nodes $0$ and $1$, respectively, primarily during periods with the highest VRE production, when the local energy demand is lower than the DA VRE forecast, and line $0-2$ is congested. This curtailment consequently increases the demand for downward reserves in these nodes.
However, the limited availability of downward reserve
at nodes $0$ and $1$, together with the congestion of line $0-2$, leads to large quantities of unmet downward reserve demand despite the high penalty for unmet downward demand.
$99.02$\% of the unmet downward demand takes place at nodes $0$ and $1$. 
At node $0$ in $65.02$\% of the hours, and at node $1$ in $6.52$\% of the hours, the uncertainty set of the balancing error $\mathcal{U}$ becomes a singleton at these nodes, i.e., only the zero balancing error is considered.
In practice, this implies that VRE at these nodes will have to self-balance in RT and cannot rely on distinct reserve providers.

\subsubsection{Sensitivity analysis on reserve demand curve penalty}
\begin{figure}[!ht]
\centerline{\includegraphics[width = \columnwidth]{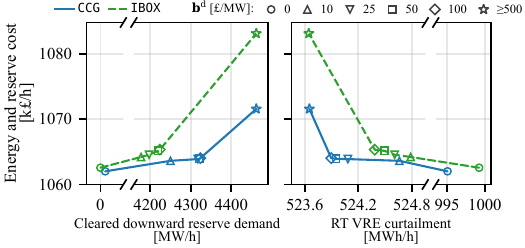}}
{\footnotesize \linespread{0.5}
Note: RT VRE curtailment is evaluated out-of-sample using hourly actual forecast errors.}
\vspace{-2mm}

\caption{Sensitivity analysis on downward reserve demand penalty $\vb^{\textrm{d}}$ (£/MW).
}
\label{sens_analysis_down_penalty}
\end{figure}

Penalties for unmet downward reserve demand are over $80$\% of the total cost (see Table \ref{da_cost}).
Since downward reserves mitigate RT VRE curtailment, the penalty for unmet downward reserve reflects wind producers' opportunity cost, which varies with out-of-market contracts.
We therefore perform a sensitivity analysis on that penalty. As Fig. \ref{sens_analysis_down_penalty} shows, the lower the downward demand penalties, the lower the amount of endogenously cleared downward reserve demand.
The benefit of \texttt{CCG} over \texttt{IBOX}, reflected in the savings of energy and reserve costs, also reduces as the penalty for unmet downward reserve demand reduces.
Below $100$ £/MW, the \texttt{IBOX} incurs higher energy and reserve costs, meets less demand for downward reserves, and has higher RT wind curtailment than \texttt{CCG}.

\section{Conclusions} \label{conclusions}

Reserve procurement methods increasingly account for network constraints as congestion can hinder reserve deliverability. 
Adaptive robust optimization has been applied for network-aware reserve procurement with decision-independent uncertainty sets for balancing errors.
This article considers uncertainty sets that depend on DA decisions regarding  cleared reserve demand and renewable curtailment.
We solve the problem using column-and-constraint generation, by reformulating the decision-dependent uncertainty sets with primitive sets.
We compare the energy and reserve schedule to a state-of-the art method, inscribed boxes, which inner approximates the real-time feasible region.
Numerical experiments on a $29$-bus system resembling the British power system illustrate how the CCG method is more cost-effective than the inscribed boxes method as it considers both positive and negative impacts of reserve activation on congestion.
The relative differences between the two methods depend on the penalty values for the scarcity of downward reserves.
Future work will extend the model to multi-interval scheduling and consider correlated forecast errors.
Further work could also provide insights into the distributional impacts of the two methods.

\bibliographystyle{IEEEtran}
\bibliography{IEEEabrv,references}

\end{document}